# Airborne forward pointing UV Rayleigh lidar for remote clear air turbulence (CAT) detection: system design and performance

Patrick Vrancken,[1,*] Martin Wirth,[1] Gerhard Ehret,[1] Hervé Barny,[2] Philippe Rondeau,[2] Henk Veerman[3]

[1] *Deutsches Zentrum für Luft- und Raumfahrt (DLR), Institut für Physik der Atmosphäre, Oberpfaffenhofen, Germany*
[2] *THALES Avionics, 25 Rue Jules Védrines, 26027 Valence Cedex, France*
[3] *Netherlands Aerospace Centre (NLR), Anthony Fokkerweg 2, 1059CM Amsterdam, The Netherlands*
*Corresponding author: patrick.vrancken@dlr.de

**A high-performance airborne UV Rayleigh lidar system was developed within the European project DELICAT. With its forward-pointing architecture it aims at demonstrating a novel detection scheme for clear air turbulence (CAT) for an aeronautics safety application. Due to its occurrence in clear and clean air at high altitudes (aviation cruise flight level), this type of turbulence evades microwave radar techniques and in most cases coherent Doppler lidar techniques. The present lidar detection technique relies on air density fluctuations measurement and is thus independent of backscatter from hydrometeors and aerosol particles. The subtle air density fluctuations caused by the turbulent air flow demand exceptionally high stability of the setup and in particular of the detection system.
This paper describes an airborne test system for the purpose of demonstrating this technology and turbulence detection method: a high-power UV Rayleigh lidar system is installed on a research aircraft in a forward-looking configuration for use in cruise flight altitudes. Flight test measurements demonstrate this unique lidar system being able to resolve air density fluctuations occurring in light-to-moderate CAT at 5 km or moderate CAT at 10 km distance. A scaling of the determined stability and noise characteristics shows that such performance is adequate for an application in commercial air transport.**

## 1. INTRODUCTION

In commercial aviation, Clear Air Turbulence (CAT) encounter is a leading cause for injuries to cabin crew and passengers and causing M$ or M€ damage per year to airlines. Non-fatal aircraft accidents and incidents are concentrated en-route where in turn turbulence encounter is the main cause for such injuries [1]. CAT encounter further yields important fatigue to aircraft structures. Thus, impact mitigation is of high interest to the aeronautics sector.

CAT is a phenomenon that is difficultly forecasted with mere provision of probability of occurrence charts over vast areas that typically cannot be fully avoided by aircraft (e.g. in the vicinity of jet streams or in the lee of mountain ridges). An in-flight forward turbulence detection of a turbulent zone ahead would allow for warning such as seat belt sign given by the flight deck. Further mitigation may consist in a slight adjustment of the flight state within the aircraft's envelope (deceleration, e.g.) or evasion maneuvers.

Turbulence in clear air, though, defies the detection by aeronautics weather radar since it relies on the backscatter of radio frequency waves on hydrometeors. Here, active optical sensing with lidar appears being the only possible and/or useful means of remotely detecting CAT [2].

In principle, airborne turbulence detection by lidar is conceivable by different methods: Wind speeds and -variations along the flight path (i.e. along the lidar line-of-sight) may readily be detected by Doppler wind lidar (DWL). Though, despite its apparent advantages, coherent DWL rely on backscatter from aerosols that are not sufficiently present at cruise flight altitudes for a thoroughly reliable and in particular long-range detection of CAT. Direct-detection Doppler techniques working on the molecular Rayleigh backscatter (with interferometric evaluation of the Doppler-shifted spectrum) necessitate a high backscattered photon number for fringe analysis. A short range implementation, for the quantification of wind speeds of gusts just ahead of an aircraft, has been demonstrated within the European AWIATOR project [3]. A long-range application as discussed here, however, imposes excessive requirements with respect to laser power or signal averaging.

Another, more photon-efficient detection principle [4] relies on tiny fluctuations of air temperature (and thus density) when air parcels undergo the up- and downwelling motion within the turbulent airflow (see Sect. 2). Thus, vertical wind speeds may be derived from air density (hence molecular backscatter) fluctuations. Vertical wind speed is also the most important parameter to know since it modifies the angle of attack of the airflow, thus directly acts on the instantaneous lift and creates the known 'bumps' and 'airholes'.

This principle of lidar turbulence detection by air density fluctuation has already been tested from ground-based lidar with some success [5].

In this paper, we report about the development, flight tests and metrological performance of an airborne Rayleigh lidar system to exploit the air density fluctuation method. This activity was performed within the European-funded FP7 DELICAT project for validating the appropriability of the method [4] and demonstrating the functional characteristics of the lidar.

The paper is organized in the following way: In the next section, the air density lidar approach is presented. Sect. 3 details and illustrates the configuration and technical details of the airborne demonstrator lidar system and its layout in the cabin of the test aircraft. Sect. 4 exemplifies the constraints of the lidar measurements with respect to meteorological conditions. It then gives a detailed evaluation of the lidar performance based on reference measurements at cruise flight altitudes. It is shown that density fluctuations as occurring in light-to-moderate CAT (or stronger) may in principle be detected by the present lidar at distances of 5 – 10 km. A scaling of the determined performance illustrates how the present lidar may extend its detection range to 25 km what would put it, from the requirements point of view, in the position of a real-world aeronautics safety application.

## 2. CAT DETECTION WITH RAYLEIGH LIDAR

Clear air turbulence results from a gravity wave (GW) field and is generated by the saturation and breaking of this field [5]. This in reality complex GW field may be engendered by orography, such as mountain waves, by convection as above thunderstorms or even shallow convection, by instabilities in stratified shear layers (Kelvin-Helmholtz-instabilities in the vicinity of jet stream borders), and commonly by a combination of these.

For the matter of lidar detection, a relationship between the vertical wind speed $w$ and the air temperature $T$ and thus density $\rho$ may be derived, as shown in reference [4]: From potential and actual temperature gradients (lapse rate) and static stability $N$ (Brunt-Väisälä-frequency), the following expression may be derived for the temperature (thus density) of an air parcel, vertically displaced by $\Delta z$:

$$\frac{\Delta \rho}{\rho} = -\frac{\Delta T}{T} = \Delta z \cdot \frac{N^2}{g} \tag{1}$$

In order to relate the relative change in density $\Delta\rho/\rho$ to the vertical wind speed $w$, we consider the critical Richardson number $Ri_c = 0.25$ below which turbulence sets in. With $Ri = N^2/S^2$ and for the shear $S = du/dz \approx 2w/\Delta z$, it follows $w = N \cdot \Delta z$ [4]. One may thus deduce:

$$\frac{\Delta \rho}{\rho} = -\frac{\Delta T}{T} = w \cdot \frac{N}{g} \tag{2}$$

With Equation (2) we obtained a simple relationship between density fluctuations and vertical velocity related to turbulent events. With typical values for $N$ of 0.01 and 0.02 rad/s for troposphere and stratosphere, respectively, and some 5 – 30 m/s vertical gust peak speed $\hat{w}$, these air density variations are very subtle, i.e. on the percent level. Considering pure molecular backscatter, they will appear as variations of the lidar signal, superimposed to the 'standard' variance arising from photon (and other) noise. For resolving the turbulence variance at useful ranges (i.e. some ten km in front of an aircraft travelling at $Ma \geq 0.8$) a high synthetic signal-to-noise ratio of, say $SNR_{av} = 100$ has to be achieved by substantial averaging of individual lidar signals. Here, these air density fluctuations can be considered 'frozen' over the considered detection/averaging time span (of some seconds to tens of seconds) compared with the characteristic rotation times of the turbulent vortices given by above values of $N$ (i.e. some five to ten minutes).

In order to most efficiently exploit the molecular backscatter $\beta_{mol}$ (in order to measure air density), a short laser wavelength is favored here, since the backscatter cross section scales with the fourth power the frequency ($\beta_{mol} \propto \nu^4$). As will be explicated in Sect. 3.A, ultra-violet radiation has further advantages, as better appropriateness to eye-safety norms and a more favorable aerosol to molecular backscatter ratio. Indeed, for aircraft en-route altitudes, the typical UV lidar backscatter ratio between aerosol and molecules $R_b - 1 = \beta_{aer}/\beta_{mol}$ amounts to less than 8×10$^{-3}$ most of the time [6, 7], which is worse with factors of 10 – 40 for NIR wavelengths. Then, we may assume for UV wavelengths:

$$\frac{\Delta \beta_{mol}}{\beta_{mol}} \cong w \cdot \frac{N}{g} \tag{3}$$

However, the above given statistical/climatological value of $R_b$ may not be expected to occur during all portions of cruise flight. Higher aerosol loads may in fact mask the desired molecular backscatter fluctuations (refer also to Sect. 4.B). Therefore, for use in an aeronautics application, a high resolution spectral (HSR) filter should be employed in order to select the spectrally narrow aerosol return and use only the molecular wings of the broad Rayleigh-Brillouin spectrum.

As mentioned above, the concept of CAT detection by air density fluctuation has been tested with ground-based lidar instrumentation from Observatoire de la Haute Provence (OHP) within the French precursor project MMEDTAC in 2008/2009 [5], even though ground-based detection (and longer-term averaging) suffers from the advection of the turbulent patch by the horizontal wind prevailing at these altitudes. The European Commission (EC) funded Sixth Framework (FP6) project FLYSAFE studied this technological alternative for use in aeronautics [8, 4]. Within the FP7 project DELICAT (2009 – 2013), such an airborne Rayleigh lidar system has been developed and flown for demonstration of its performance and proof of concept. The central requirement formulated for the DELICAT project was to measure a 1 % air density fluctuation at 5 km distance in cruise-flight altitude, corresponding to moderate CAT.

Within the project, the lidar system was integrated in a research aircraft in forward-pointing arrangement. In this configuration, the aircraft passes through the farthest lidar-probed zones (recorded detection range: 15 km) after less than 1:30 - 2:00 min (depending on altitude / flight speed), provided the absence of horizontal wind shear during the respective flight portion. With the assumption of frozen turbulence, the aircraft itself acts as a 'truth' sensor for the turbulence. An inertial reference system (IRS) delivers high-resolution acceleration for all six axes. This macroscopic aerodynamic turbulence sensor, though, is rather elaborate to exploit in terms of turbulent vertical velocity, necessitating complex aircraft and also turbulence models. Thus, the aircraft was also equipped with a fast total air temperature probe also delivering a low-pass filtered turbulence 'truth' according to Equation 2. The respective prevailing larger-scale Brunt-Väisälä frequency $N$ may be interpolated from temperature measurements between ascents and descents next to a turbulence encounter or taken from numerical weather analysis.

## 3. LIDAR SYSTEM

A synopsis of the developed airborne lidar system is depicted in Fig. 1. A high power laser transmitter sends short pulses on a beam steering device. This device compensates the attitude and movements of the aircraft and ensures the horizontal projection of the laser beam on the ahead-lying flightpath. The backscattered light takes the same way, is collected by a telescope, filtered and projected onto a set of detectors. These elements, together with their performance, are described in the following sections A through F.

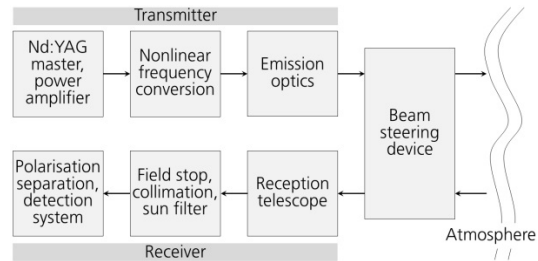

Fig. 1. Synopsis of the DELICAT lidar system

The lidar system is mounted within a stiff rack structure which is adapted to the NLR research aircraft PH-LAB, a modified Cessna Citation 2 aircraft. To allow the horizontal projection of the laser beam onto the flight path, it is equipped with a special aerodynamically optimized fairing under which is located the forward bending mirror. Fig. 2 shows photographs of this carbon-fiber reinforced plastic structure on the starboard side of the aircraft and a view on the lidar system inside the cabin.

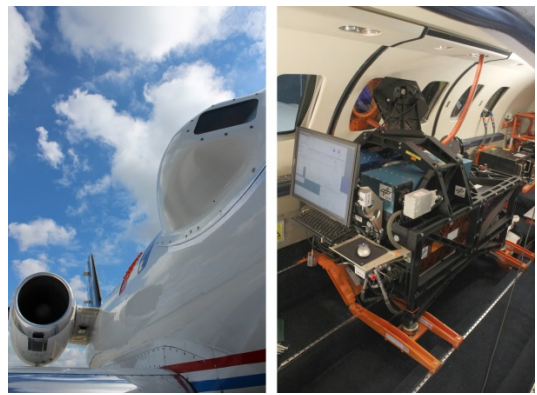

Fig. 2. Left: View on the outside starboard fuselage with the fairing for laser beam transmission and reception. Right: Complete lidar and beam steering system with operator interface.

### A. Transmitter

For a given realistically achievable laser power, this application calls for high pulse energies with low repetition rate rather than low pulse energy with high rate since the averaged $SNR_{av}$ scales linearly with the former and by the square-root for the latter. In the same logic, regarding the choice of an appropriate wavelength, one may consider maximizing the following simple figure-of-merit:

$$F.o.M. = \eta_{c_v} \cdot R_{\beta_\lambda} \cdot Q.E. \qquad (4)$$

where $\eta_{c_v}$ is the laser frequency conversion efficiency, $R_{\beta_\lambda}$ the ratio of Rayleigh backscatter coefficients at different wavelengths, and $Q.E.$ the quantum efficiency of a detector at a considered wavelength. The following table shows typical achievable values of this $F.o.M.$ for the three fundamentals of the well-proven solid-state Nd:YAG laser for two different detector types, photomultiplier tubes (PMT) and avalanche photodiodes (APD):

Table 1: Parameters for Rayleigh lidar figure-of-merit

| $l$ / nm | $\eta_{c_v}$ | $R_{\beta_\lambda}$ | Q.E.PMT | Q.E.APD | F.o.M.PMT | F.o.M.APD |
|---|---|---|---|---|---|---|
| 1064 | 1 | $(1/3)^4$ | $10^{-3}$ | 0.8 | $1.2 \cdot 10^{-5}$ | $9.6 \cdot 10^{-3}$ |
| 532 | 0.55 | $(2/3)^4$ | 0.2 | 0.75 | $2.2 \cdot 10^{-2}$ | $8.3 \cdot 10^{-2}$ |
| 355 | 0.3 | 1 | 0.3 | 0.25 | $9 \cdot 10^{-2}$ | $7.5 \cdot 10^{-2}$ |

This exercise shows the principal equivalency of using the green second-harmonic and the UV third-harmonic. However, since the visible harmonic imposes more serious eye-safety considerations than the invisible third harmonic (see Sect. 3. G.), a UV laser source based on non-linear frequency conversion is favored.

As a source, the several times flight-proven pump laser of the DLR WALES lidar was chosen. It is thoroughly described in [9], some details are given in the following (cf. Fig. 3). The laser is of the MOPA (master oscillator, power amplifier) design, with a monolithic intrinsically single-mode running Nd:YAG master resonator. This diode-pumped NPRO (non-planar ring oscillator) laser emits about 150 mW of infrared laser pulses with a length (FWHM) of 8 ns at a rate of 4 kHz. Its frequency may be tuned and modulated both by temperature and mechanical stress. These techniques are used for locking it to a molecular reference. The NPRO oscillator is stress-modulated with a sine wave that is in phase with the reference signal (see below) for the subsequent power amplifiers. A small part of the IR radiation is directly frequency doubled (SHG) and fed through an Iodine vapor absorption cell generating a signal on a photodetector (PD). This is used in a lock-in technique to stabilize the laser to the absorption line center yielding an absolute frequency stability below 1 MHz and 300 kHz on short time scales (< 1 min). Note that this frequency locking is not required in the current setup but was used nonetheless. The laser is thus apt for use in a possible future setup including a high spectral resolution lidar (HSRL) setup (as addressed in Sect. 2).

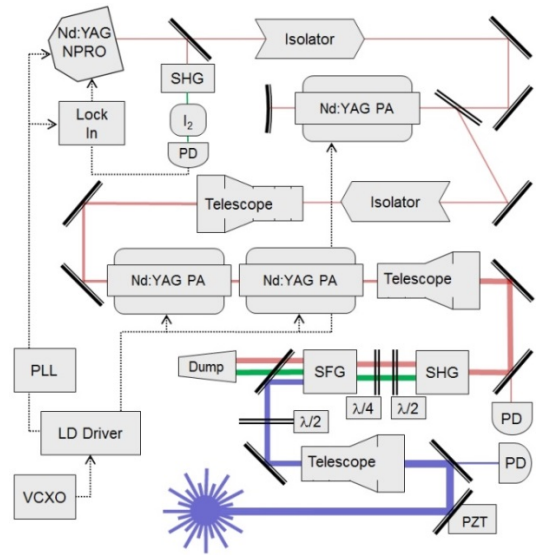

Fig. 3. Schematic optical layout of the transmitter. See text and [9] for details and acronyms. Only main elements are shown here.

The absolute repetition rate reference (at 100 Hz) of the MOPA setup is imposed by the driver current cycle of the power amplifier (PA) stages. The residual timing jitter of the passively Q-switch generated pulses is less than 0.5 µs (at 1 σ) leading to a very low pulse-to-pulse power variation (see below). The amplifier setup is composed of a small-signal double-pass amplifier and two single-pass main amplifiers. With a combined gain of 40 dB, the laser thus delivers pulse energies of up to 400 mJ. Measurements of the fundamental at this power level yield a beam quality of $M^2$ = 1.5. A photodiode (PD) registers the outgoing pulses what is used for the triggering of the data acquisition (see Sect. 3.D).

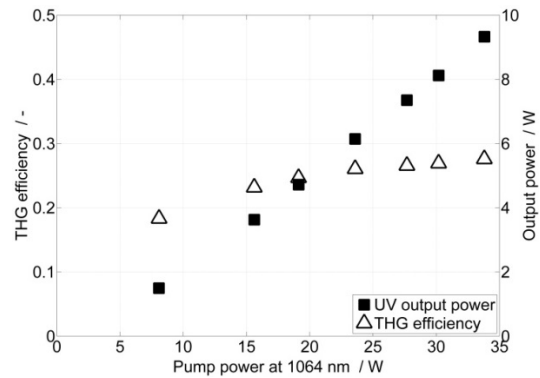

Fig. 4. Third-harmonic generation efficiency and output power

The infrared radiation is then fed into a KTP crystal for second-harmonic generation (SHG) in type II configuration ($oe \rightarrow o$). The phase-matching is achieved coarsely by angle tuning and finely by temperature tuning. For this purpose, the KTP crystal is heated to approx. 80°C and the whole setup is accommodated in a heat-insulated compartment. Thus, at full laser power, a SHG conversion efficiency of up to 55 % is achieved. Upon SHG, both the fundamental as well as the second harmonic feature a certain elliptic polarization. Therefore, a set of two-wavelength zero-order waveplates with $\lambda/4$ and $\lambda/2$ delay are employed to adjust the two harmonics' polarization state to the linear state. Then the beams are fed into a BBO crystal for sum frequency generation (SFG). Here, the phase matching is achieved by angular tuning within a Piezo-driven two-axes mount. Together, a third-harmonic generation (THG) efficiency of up to 30 % is achieved. Fig. 4 shows the THG efficiency dependency over the incoming laser power and the respective attained power at 355 nm.

The UV radiation is then led over a first highly dichroic mirror; the transmitted infrared and green portions are fed into a beam dump. The UV part is directed through a motorized zero-order $\lambda/2$ waveplate in order to rotate its polarization for optimizing the transmission losses over the skew arrangement of many (eight) mirrors to follow in the subsequent optical setup. Some leakage of the UV light behind a second dichroic mirror is focused on a fast PIN photodetector; the generated signal is fed through a sample and hold circuit and digitized. This monitoring of the laser pulse energy of every emitted pulse is stored alongside the lidar information (see Sect. 3.D). Fig. 5 shows a comparison of this internal laser pulse energy measurement with an external (continuous) pyro-electric sensor.

The measurement reveals the linearity of this pulse energy measurement process and the low pulse-to-pulse energy jitter of less than 0.5 % which have been measured quite equally internally (dispersion 0.1 to 0.4 mJ) and by the reference (dispersion 40 mW). For the internal pulse measurement this includes the pulse-to-pulse energy fluctuation originating in the passive Q-switch pulse timing jitter, subsequent fluctuation in laser amplification which is then further deteriorated by the non-linear efficiency of the THG process. This implies an excellent precision of this digital pulse energy measurement.

The main high-power UV laser beam is expanded by another Galilean telescope to a diameter of 13 mm. Further, its divergence is adjusted to a value $\Theta$ = 150 µrad. The beam quality factor in the UV was determined to $M^2$ = 4.3 which is in good agreement with theory on beam quality degradation with nonlinear frequency conversion: $1 \leq M^2_{3\nu}/M^2_{1\nu} \leq 3$, depending on conversion efficiency [10].

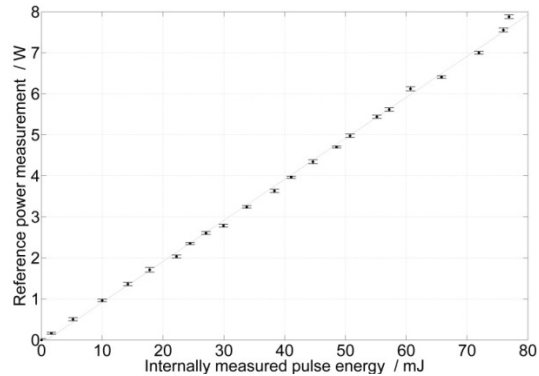

Fig. 5. UV laser pulse energy and comparative power measurement with external reference. The line shows a linear regression with mean residuals of 60 mW. These originate mainly in the pyroelectric detector which features a dispersion of around 40 mW (vertical error bars).

The beam is then guided through a shutter device that allows the lidar operator in the aircraft cabin and the flight deck to remotely block the laser output without interrupting laser operation. Another dichroic mirror directs the UV beam into the subsequent beam feed.

Opto-mechanically, the whole laser and harmonic generation assemblies are based on standard laboratory holders and some custom parts, both from aluminum and steel alloys. A flight-proven compact rugged vibration- and ambient condition resilient design is followed. Fig. 7 shows a photo of the all-self-contained transmitter. The upper part (visible in the image) contains all the optical elements referred to above (as in Fig. 3), with the master oscillator to the far right rear and the three greenish amplifiers in the red compartment and the THG in the magenta module. The THG module may be exchanged to any other OPO (optical parametric oscillator) of the WALES lidar series [9, 11].

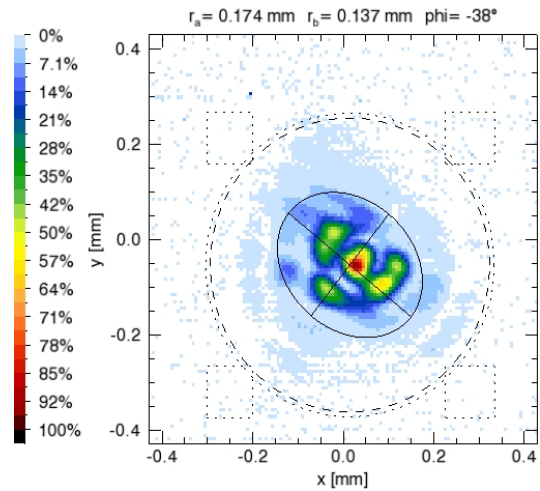

Fig. 6. Far-field beam pattern of the third harmonic of the laser

The lower half of the transmitter body contains all electronics, in particular the DC/DC converters and diode drivers of the laser power amplifier stages. The intermediate level of the IR pump laser features a water cooling circuit directly feeding the pump chambers with de-ionized water. The overall power consumption at full optical output power is 800 W. The chief part of this power is evacuated as heat by the water cooling circuit. The heat is then exchanged to another cooling circuit based on some aeronautics coolant oil. This circuit transports the heat charge it to an external cooling plate which is installed in a cabin windows aperture. The transmitter measures 935 × 412 × 257 mm³ including the THG compartment with a total mass of 106 kg. It is integrated in a stiff rack (see Sect. 3.F) with the main other optical subsystems like receiver and beam guidance device.

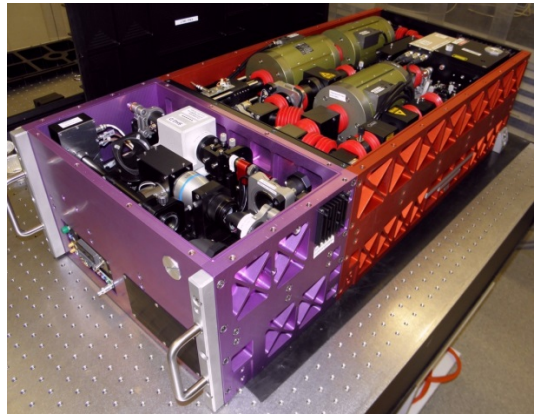

Fig. 7. Photo of the (opened) transmitter, see text for details.

**B. Laser beam guidance system**

The lidar system is arranged in a classical monostatic arrangement, with the laser beam being transmitted over a mirror superposed to the secondary mirror of the receiver telescope (receiver see next section). To this end, the laser beam is guided through a tubular system from the transmitter exit to this transmit mirror. From here on, the optical transmit and receive paths are common. The tubular beam feed comprises a piezo-electrically motorized mirror. This serves to fine-tune the transmit beam into the receiver field-of-view (FOV) which is defined by the telescope optics and field stop.

The common transmit/receive path is guided over a complex arrangement of mirrors and windows out of the aircraft cabin where it is folded forward onto the flight path.

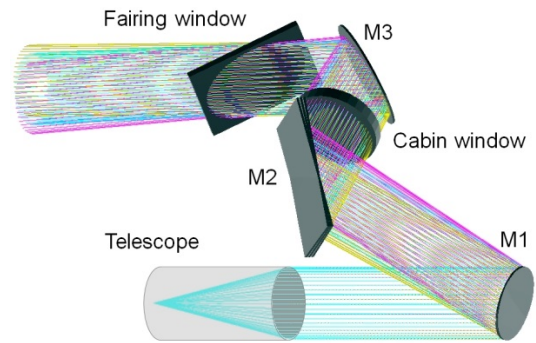

Fig. 8. Schematic of the common transmit/receive beam guidance system (ray-tracing with ZEMAX). Shown are three different angles of the beam outside the aircraft which are controlled by movable mirrors M1 and M2. See text for details.

During steady flight (high altitude cruise) where the CAT-measurements should take place, the aircraft is subjected to following movements: Change of aerodynamic angle of attack due to different altitudes, speeds and change of mass and center of gravity (due to fuel consumption); low frequency residual aircraft dynamic modes like phugoid and Dutch roll that may not fully be damped by autopilot or other means; and movements due to the light background turbulence. In order to ensure probing the same air volume at distances of 5 to 15 km within reasonable limits, these movements have to be compensated by steering the transmit/receive beam accordingly. As a primary constraint, the front window of the aircraft aerodynamic fairing (see Fig. 2) and its geometry with respect to the cabin (and its windows) are predetermined.

The fairing front window's footprint allows for an effective receive beam diameter of only 140 mm, hence it was chosen to make its center an invariant of the beam steering system. For development convenience it was further chosen to implement the forward folding mirror within the fairing as fixed since it is unpressurized and thus undergoes important temperature gradients and absolute values of as low as -50°C within any flight. Last, due to the limited available area, also the optical surfaces are implemented common, i.e. without physical separation (such as tubes and window frames) between transmit and receive areas. Anticipating Sect. 3.C, it was determined that the laser scattering in particular from the windows does not have a negative effect on the analogue detection devices. A ray-tracing schematic of the implemented beam guidance system is given in Fig. 8 while the movable mirrors may be discerned on the CAD-image of Fig. 11.

Following the transmitted photons from the transmit folding mirror in front of the receiver telescope, the beam is first bent by two two-axes movable mirrors M1 and M2. Both must accommodate the beam walk due to the invariant point being around a meter further ahead. After M2, the beam passes through a first window: This 'cabin' window separates the pressurized cabin environment from the ambient air. It is a 20 mm thick 200 mm diameter fused silica window in a special holder which is integrated in a plate within the frame of a standard cabin window aperture. Extensive strength calculations and tests have been performed (see Sect. 3.F) for the airworthiness certification of this window/frame ensemble. On the outside of the cabin, still inside the fairing, the beam hits M3 that bends it forward. Mirror M3 is fastened in a holder that is also attached to the previous window plate. Last, the beam passes through the fairing front window that is directly clamped into the carbon-fiber reinforced structure of the aerodynamic fairing. This, a development of the past EC FP5 I–Wake project (DLR coherent Doppler lidar for wake vortex measurement [12]) has been upgraded (i.e. reinforced) to meet the aerodynamic load requirements of the high-altitude flights in CAT research. All windows and mirrors were budgeted to fulfil a low net wavefront error.

The absolute dynamic of the system is 2.5° in pitch and 1° in yaw. Based on a coordinate transformation matrix (which is determined by calibration) the mirrors' motorizations are steered by a software that reads the aircraft attitude information (pitch from the laser-gyroscopic inertial reference system (IRS), yaw as side-slip-angle from a vane on an external noseboom). Based on these data, it continuously performs an aircraft movement extrapolation and calculates the necessary actuator accelerations, also based on their actual state. The typical overall dynamic variations have been determined from previous flight data and are covered by the system which yields a bandwidth of 1 Hz (at 0.25° amplitude). Tests of the system with simulated input data have shown a standard deviation of < 0.05° for both axes which complies with the requirements. Fig. 9 shows a position plot of this test.

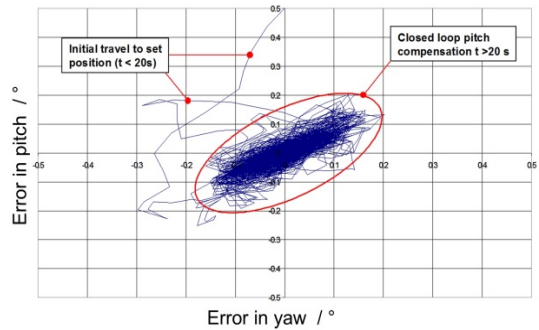

Fig. 9. Dynamic behavior of the beam steering system.

## C. Receiver system

As pointed out above, the receiver/transmit architecture of the lidar system is of the monostatic type. This receiver is a separate module which is easily interchangeable on the lidar rack platform. Its layout is likewise modular, offering the possibility to implement different versions. Common is the $f/5$ Newtonian telescope with 150 mm primary mirror (a 140 mm masked aperture is used). Both secondary and coaxial transmit mirrors (with mounts) are designed to yield a small obstruction ratio of 12 %. The front optics include a field stop of 400 to 800 μm diameter in the tests for allowing for some inadvertent movement of the laser beam within the field of view (FOV), an aspheric beam collimation assembly and a narrow interference filter. This sun filter yields a width of 0.5 nm (FWHM) with an 88 % peak transmission and an out-of-band blocking of OD4.

For these first flight tests, the receiver was laid out as a simple backscatter receiver with an additional depolarization channel. This was implemented in order to easier recognize, albeit not exhaustively, airspace areas containing aerosol that would contaminate the desirable pure molecular backscatter return. This concept has been proven with some success in ground-based measurements during the French precursor project MMEDTAC [5]. Refer to Sect. 3.G for details on the operational implications regarding aerosol.

The depolarization assembly consists of a set of two waveplates ($\lambda/2$ and $\lambda/4$) and a polarisation beam splitter. Hence, the system features a parallel and a perpendicular channel. Both collimated beams are imaged with lenses into the central areas of two UV-optimized PMTs (for insuring misalignment insensitivity). Their bialkali photocathodes feature a responsitivity of around 58 mA/W at 355 nm. The PMT-voltage may be driven to yield a gain from $5\cdot10^3$ to $3\cdot10^6$ and has been operated at 500 - 580 V (gain: $2.5\cdot10^4$ - $7\cdot10^4$). The photomultipliers are operated within self-contained modules that are also derived from the WALES lidar [9]. The signal currents are amplified by an especially designed transimpedance amplification circuitry that ensures a low noise output (measured to 0.3 fW/√Hz including amplifier and digitizer contributions) and a 100 ns (FWHM) pulse response. The detection module further contains the high-voltage supply circuitry and the analog-to-digital converters (ADCs) with 12 bit resolution. The digitization rate is programmable, 30 MHz have been chosen for the present tests. All these elements are well shielded in a tight housing in order to reduce to a minimum the electromagnetic interference from laser power supplies and other high-frequency sources such as computers and aircraft radio systems. The digitized lidar signals are routed to a central data acquisition and control unit (see next section).

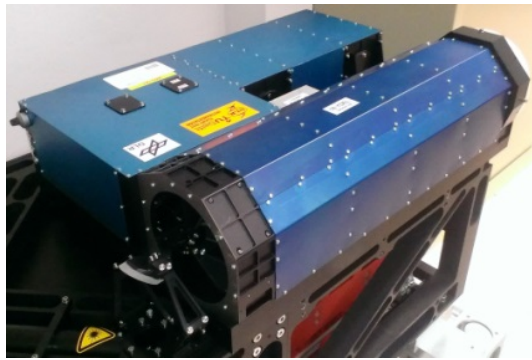

Fig. 10: Photo of the receiver system, see text for details

The calibration of the depolarization unit was performed during a high-altitude calibration flight leg with presumably insignificant aerosol content (FL380 over the British islands on Aug. 6, 2013; details on the flight campaign see Sect. 4.A). Here, the polarization assembly was optimized by minimizing the signal on the perpendicular channel by rotation of the waveplates. Thus, these polarizing elements define the system's reference polarization state which is aligned with respect to the emitted one. The minimized measured depolarization was determined by commuting the channels with the help of the waveplates, making thus it independent of the channels' individual (detector) gain. The mean value was thus determined to $\delta_m = 1.4\ \%$. According to [13], with some simplifications owing to the fact that the same detectors are used for both channels and neglecting effects such as attenuation, one may determine:

$$\delta^S = \frac{\delta_m - \delta^V}{\delta_m + 1} \qquad (5)$$

with $\delta^V$ designating the volume and $\delta^S$ the system depolarization, including all elements from laser emission and transmission over the optics up to the depolarization assembly. The volume depolarization due to rotational Raman lines passing through the narrow sun filter may be determined in the following way: Since the shift of these lines is constant on a wavenumber (or frequency) scale and their relative intensities approximately independent of the laser frequency, the results of the very helpful publication [13] may be transformed from 532 nm to 355 nm by simple scaling. For such narrow filters as ours with 0.5 nm width, the temperature dependence is negligible and thus one determines from [13, Table 2] the volume depolarization to $\delta^V_{355nm} = 4.7 \cdot 10^{-3}$. And thus for the lidar system depolarization:

$\delta^S = 0.9\%$

what is a rather good value and allows for reliable depolarization measurements.

### D. Data acquisition and control unit

The lidar signal acquisition and control is performed by a single device which is a strip-down version of the WALES unit. Besides a computer and some DC power supplies it contains: a commercial custom-built digital IO-board for the signal acquisition from the detection modules; a GPS receiver for time stamping of the laser shots; an ARINC aeronautical standard bus board for storing aircraft avionics and experimental data; and a variety of self-developed electronics as the lock-in amplifier for wavelength stabilization, interfaces for the Piezo amplifiers and timing electronics for the different trigger signals.

The lidar signal generation (i.e. digitization of PMT output) runs continuously while the data acquisition (i.e. storage) is triggered by the transmitter clock oscillator – it is started 100 μs before a laser pulse is emitted to allow for the some μs Q-switch jitter. For the present application, each individual lidar signal is stored.

**Table 2: Lidar system parameters**

| Parameter | Value |
|---|---|
| Repetition rate | 100 Hz |
| Laser pulse energy (IR) | 310 mJ |
| Laser pulse energy (VIS) | 230 mJ |
| Laser pulse energy (UV) | 85 mJ |
| Line width (IR) | 54 Mhz |
| Frequency stability (IR) | ≤ 1 MHz |
| Pulse length | 8 ns |
| Beam quality $M^2$ (UV) | 4.3 |
| Beam divergence | 150 μrad |
| Telescope diameter | 140 mm |
| Field of view | 530 μrad |
| Sun filter bandwidth (FWHM) | 0.5 nm |
| System depolarisation | 0.9 % |
| Detector NEP | 0.25 fW/√Hz |
| Analog/digital converter | 14 bit |
| Sampling rate | 30 MHz (5 m) |
| Data storage | Shot-by-shot |
| Beam steering dynamic | 1 Hz (sine) |
| Pointing accuracy | 0.1° |

A dedicated software environment allows the operator to follow in real-time the lidar operation. It permits controlling the main parameters of the transmitter (laser frequency and –locking, THG parameters, shutter, transmission beam angles, etc.) and the receiver (PMT voltages). The software environment provides (nearly) shot-to-shot quicklooks of the acquired signal and basic real-time signal conditioning functions (averaging, scaling, etc.) in order to efficiently monitor the correct operation.

Physically, the data acquisition and control unit is housed in a 4U 19" drawer in a separate rack, while the screen and input device are attached to the lidar rack (see Fig. 11). An operator seat is arranged directly in front of it.

The system parameters of the overall lidar system are summarized in Table 2.

### E. Test aircraft integration and certification

The modular design of the lidar system is in part due to the fact that the subsystems have been developed by two (initially three) project partners (DLR for the lidar, THALES Avionics for the beam steering unit), all interfacing with the NLR aircraft. The other reason is the deliberate decision for flexibility for future modifications and extensions. The different modules are mounted onto a rigid rack structure. For airworthiness, this structure has to be designed to the extreme loads (emergency landing case) defined by EASA CS 25 (i.e. JAR/FAR Part 25) [16], that is 9.0 g, 3.0 g and 6.0 g for the three axes forward, side, downwards. This exercise results in the likewise modular rack structure depicted in Fig. 2 and Fig. 11. A main frame embraces the transmitter and serves as a basis for the receiver, the first beam steering mirror, an upper structure for the second beam steering mirror and screen/keyboard.

The whole is mounted on a lower structure that interfaces (via appropriate shock mounts) another set of structures stretching three parallel seat-rails (across the cabin aisle). The remaining parts of the lidar system, i.e. the front-facing mirror (and cabin window) are attached to the cabin window plate, meaning mechanically separated from the primary lidar structure. This implies the consequent possibility of a misalignment (of lidar beam to flight path) due to differential movement of fuselage and cabin-internal structure (such as seat rail), which is due to cabin expansion and application of forces from wing or landing gear. In dedicated tests prior to the lidar flight campaign, these movements were found to be present mainly in the rotational axis (around aircraft longitudinal axis) and generally fitting in the beam steering envelope of ± 0.1°.

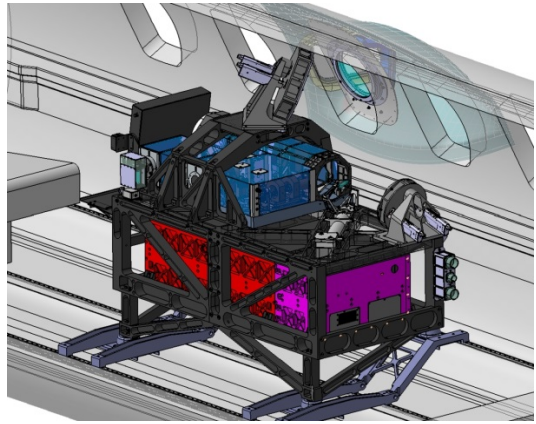

Fig. 11. CAD view of the lidar system installed in the cabin of the PH–LAB aircraft (from CATIA)

The modified aircraft with the whole test ensemble including lidar system and all other aircraft modifications such as nose boom (with side slip vane, see Sect. 3. B) and aircraft experimental systems (IRS etc.) received a supplemental type certificate (STC) from Dutch CAA in July, 2013.

**F. Eye safety**

Another key issue for certification and operation within airspace is the laser eye safety. Different aspects have to be covered here: first, internal to the aircraft, concerning operators and aircrew; second, outside the aircraft regarding possible scatter of the laser beam and pick-up by the crew; and last, outside the aircraft, within regular airspace regarding other aircraft.

Within the aircraft cabin, for the protection of the lidar and test system operators, a Class-1 environment according to IEC 60825–1 [17] is established by an opaque enclosure of the lidar system made of an appropriate light-tight fabric.

Regarding the pick-up of laser beam scatter by the flight crew (the laser beam passes approximately 0.5 m to the right of the Flight Officer's (FO) seat), a series of highly worst case assumptions has been run through. The most efficient inadvertent scattering mechanism (in flight) may be constituted by very dense clouds. With conservative use of backscatter values from dense clouds [7, 18, 19, 20], it can be shown that the fluence of the produced scattered pulse (integrated from 0.5 m to infinity for an infinite, uniform cloud) remains a factor $2 \cdot 10^4$ below the maximum permissible exposure (*MPE*) derived from the norm IEC 60825-1 [17, Table A.1]. A fully hypothetic long-haul flight of 10 h, all in dense clouds, would still leave a security factor of 500 between the irradiation and any critical level (of repeated exposure).

The main concern of authorities is the protection of other airspace users from this high-power laser beam. Given the high relative speeds of aircraft the multiple pulse considerations do only apply for two aircraft with absolute opposite course. For any other course, the single pulse *MPE* pertains, resulting in a 'hazardous' distance of 333 m, usually called the NOHD (Nominal Ocular Hazard Distance). We desist from assessing the absolute probability of some encounter aircraft and our research aircraft finding themselves in the proper geometric and temporal (given the laser duty cycle of $< 10^{-6}$) configuration - but given the physical/mechanical limitations of the beam architecture or the aircraft operational pitch limits, even when in the closest parallel flight levels (1,000 ft apart), still comfortable security factors of 20 and 3, respectively, result. The flight with opposite courses, which would necessitate taking into account repeated laser pulse exposure, is ruled out by the measures of air traffic management.

In summary, it can be stated that the use of a high-power laser beam (at least) in the described configuration may be regarded as completely eye-safe for operators, air and cabin crew as for other aircraft in the surrounding airspace.

## 4. TEST AND PERFORMANCE EVALUATION

**A. Flight test campaign and measurements**

Before the flight tests planned within the DELICAT project, the lidar has been completely integrated as per Fig. 11 (omitting the beam steering and aircraft mirrors and windows) in the DLR premises at Oberpfaffenhofen. The lidar transmit/receive path were guided over a 400 mm UV mirror and fed vertically through a roof opening in one of DLR's lidar labs. Measurements during the winter months of 2013 allowed validating the lidar functionality and optimizing optics, electronics and data handling. The reliable referencing of the lidar internal alignment (transmit vs. receive) prior to aircraft integration and flight test was particularly important since the lidar may not be fully operated from ground once integrated in the aircraft cabin due to eye safety constraints on the home airport Schiphol, Netherlands.

Upon validation at DLR, the lidar was disassembled and transported to NLR at Schiphol-Amsterdam for integration with the test aircraft PH-LAB in May 2013. A series of alignment sessions allowed guaranteeing the lidar pointing on the projected flight path with a high level of confidence, including: the alignment of the aircraft nose-boom (measuring side-slip) w.r.t. to the IRS; the alignment of the nose-boom w.r.t. to the aircraft fuselage; the pointing of the beam steering (with low power laser beam) w.r.t. the aircraft fuselage; as pointed out above, the lidar-internal transmit/receive alignment could only be roughly checked with a collimated beam alignment system without active lidar operation.

After the issue of a "Supplemental Type Certificate" (STC) from Dutch Civil Aviation Authority to the heavily modified PH-LAB, the flight test campaign was executed from July 17 to August 12, 2013. It contained eleven Europe-continental flights from Schiphol airport, summing up to more than 30 flight hours during which the lidar was nominally operated over 15 hours. Despite the meticulous planning and preparation of each flight (see next section) for attaining CAT-infested areas, the yield in turbulence measurements was rather low. This was mainly due to the prevailing weather conditions, a suite of very stable High systems during the 2013 summer, a season that was not aimed for in the project (preferring rather winter) but could not be chosen due to previous delays in the project. The present publication aims at the demonstration of the system performance itself (Sect. 4.D) rather than on the turbulence measurements.

**B. Operational aspects: Meteorology and Aerosol**

As pointed out in the Introduction, CAT (of a certain, aircraft-perceivable strength) is a quite rare phenomenon which evades a precise forecast. Being parametrized in numerical weather prediction (NWP) models, a forecast [24, e.g.] is usually performed by exploiting different CAT indices, such as Richardson number, turbulent kinetic energy $TKE$, Ellrod- or Colson-Panofsky index (to cite some amongst over twenty). Within DELICAT, project partners University of Warsaw and MétéoFrance provided such exploitation, both by human aeronautical synoptic forecasters and an automated combined index approach. The forecast was performed several times per day depending on the prevailing conditions and flight planning.

The met partners further provided forecast on cloud (Cirrus) cover and expected aerosol load. The latter is particularly important due to the current lidar receiver design without any high spectral resolution (HSR) filter for aerosol signal rejection. The implemented depolarization channel merely facilitates the identification of lidar measurements with elevated aerosol backscatter (and thus its rejection in the analysis), and this only for depolarizing aerosols.

Therefore, it is important to design the mission such that aerosol backscatter does not significantly hamper the set goals. As pointed out in the Introduction, we assume being able to resolve the density (i.e. Rayleigh backscatter) fluctuation originating in turbulence (in clear air, devoid of aerosols) at some distance when attaining a (synthetic) $SNR_\Sigma \geq 100$, i.e. by averaging a number $N_\Sigma$ of data points, temporally and/or spatially. Then we consider an additional signal $S_{Aer}$ originating in aerosol backscatter, with the proportion $S_{Aer} = k \cdot S_{Ray}$. One may suppose that the aerosol signal remains constant over the averaged domain (as in [5]). While this supposition should hold true for temporal averaging, it may be more precarious for spatial averaging (as in Sect. 4.D) for the aerosol concentration possibly being variable - in particular when originally a vertical gradient was present (before turbulence setting in). However, taking this assumption granted, the aerosol signal may be subtracted, solely its noise remains. Thus, in order to attain the same $SNR_\Sigma$ as for pure Rayleigh backscatter, the number of averaged measurements $N_\Sigma$ has to be increased by the same amount $k$. For risk reduction it was chosen $k < 10\%$ what results in a maximally permissible backscatter ratio $R_{b_{max}} = \frac{\beta_{Mol} + \beta_{Aer}}{\beta_{Mol}} = 1.1$. Statistical analysis of lidar measurements [6, 7] over the Atlantic, for instance, show that the aerosol backscatter in the UV stays well below this level on average (Fig. 12: Median and upper quartile).

This backscatter compendium however emphasizes the "historically clean period, 1988-1990", why these values "likely provide a background level". Further, a perceivable proportion shows high backscatter approaching the molecular level.

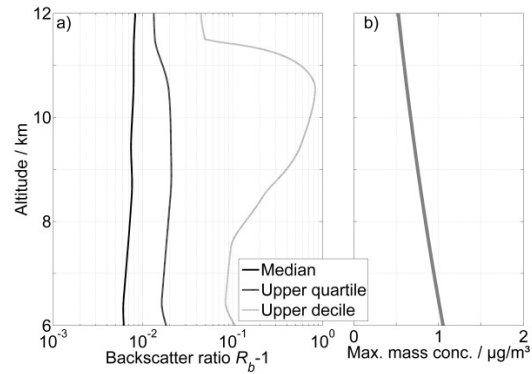

Fig. 12. a) Backscatter ratio at 355 nm derived from [6, 7], actually giving a "median of median values" from 80 flights in six different areas and seasons of the Atlantic ocean region, shown at the flight altitudes relevant for DELICAT. The graph also gives the 25% and 10% quantiles for the unfavorable case. b) Determined maximum permissible aerosol mass concentration for the flight tests.

For these reasons the reliance on a climatology was not followed in the DELICAT project; instead, a direct forecast of the aerosol occurrence was aimed. For that purpose, MétéoFrance provided outputs from the chemistry and transport model MOCAGE [25] that was forced by the ARPEGE global NWP model. MOCAGE provides mass concentrations for aerosols covering accumulation and coarse particle modes for different altitude (pressure) layers.

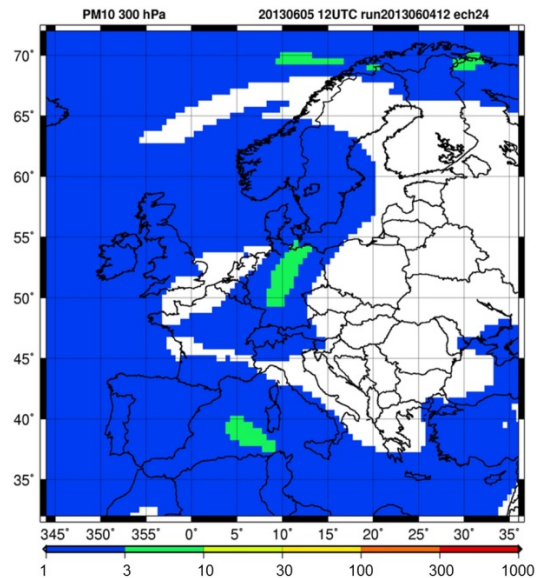

Fig. 13. Sample of MOCAGE aerosol forecast with color-coded mass concentration $M$ in µg/m$^3$ at 300 hPa.

For a derivation of aerosol backscatter based on these forecasts, a complete aerosol microphysics and trajectory analysis had however to be renounced for obvious complexity reasons.

Instead, a worst-case approach was followed. The software OPAC (Optical Properties of Aerosols and Clouds) allows calculating optical properties of all kinds of aerosol types and mixtures [26]. The optically most efficient aerosol type in terms of scatter to mass ratio occurs for low density, high real part of index of refraction and imaginary part equaling zero. As a worst case for our application in upper atmospheric layers may thus be considered a dry water-soluble "WASO" aerosol [27], with a density of 1 g/cm$^3$ and real part of refractive index $Re(n)$ = 1.52 which leads to a maximum [28] for the ratio of backscatter coefficient and mass concentration, betimes called the (particle) mass scattering coefficient [29]:

$$MSC_{max} = \left(\beta_{Aer}/M_{Aer}\right)_{max} = 0.4 \ m^2/sr \cdot g \tag{6}$$

A more moderate ratio may be derived when assuming slightly moist WASO ($Re(n)$ = 1.40) leading to the half of this value.

Comparative values to this computed hypothetical case obtained from field measurements exhibit a wide range depending on aerosol density and mode. Table 4 gives some arbitrarily chosen samples for the mass scattering coefficient $MSC$ for diverse aerosol types and analyzed altitudes (mainly boundary layer). The cases are Saharan mineral dust (MD) advected over Germany and Romania, volcanic ash (VA) from the Eyjafjalla and Pinatubo eruptions, marine boundary layer (MBL), biomass burning (BM) in Amazonia and mixed aerosol (MIX) over Greece.

**Table 3: Mass scattering coefficient for different aerosol types**

| Case | $MSC$ / m$^2$/sr·g | Reference |
|---|---|---|
| MD Germany | 0.008 | [30] |
| MD Romania | 0.024 | [31] |
| VA Eyjafjalla | 0.04 | [30] |
| VA Pinatubo | 0.15 | [32] |
| MBL Korea | 0.02 | [33] |
| BM Amazonia | 1.3 | [29] |
| MIX Greece | 0.05 | [34] |

Here, the backscatter coefficient has been transformed from the original ones (visible wavelengths) to UV by applying the scaling law reported in [7], even if this may not take account of the diverse microphysics.

The Amazonian case clearly exceeds the determined worst-case value since it originates in optically very active "fine mode dominated" biomass burning aerosol. Conjecturing that such fine mode aerosol is of too short lifetime to be found in the UTLS (upper troposphere – lower stratosphere) aimed at in DELICAT, we are confident that the above given $MSC_{max}$ actually represents a conservative and worst-case value for the flight planning, with a viable security margin.

The flight planning thus considered the aimed flight altitude and permissible aerosol mass concentration determined from $MSC_{max}$ and $R_{b_{max}}$, as given in Fig. 12 b). The MOCAGE forecast was thus used to identify the areas with a mass concentration of approximately < 1 µg/m$^3$, thus with the motto "aim the white (at the limit blue) areas!" Actually, in practice it was not that hard to do this since aerosol occurrence luckily fell short of apprehension.

Actually, Cirrus clouds represented the main limitation since they often coincided with the turbulence-forecasted regions. On the other hand, Cirrus may be well detected by the depolarization channel of the receiving system.

**C. Lidar signal evaluation**

For checking the performance of the lidar system, among the 15 h dataset three flight segments have been selected for a deeper performance analysis. As mentioned above, the encountered turbulent events have been rather weak, and well below the detection threshold as will be detailed in the next section. For this reason we will analyze here 'zero' measurements with presumably low aerosol or Cirrus content and without turbulence. These 'zero' reference measurements have been carried out in the local vicinity of some very light turbulent events, though, which has actually been their selection criterion. With the dispersion over the whole campaign period and different flight altitudes, we assume that the following findings are representative of the whole campaign dataset. The following table summarizes the details of these reference measurements:

**Table 4: Reference flight legs with 'zero' measurements**

| # | Date | Location | Altitude | Time span |
|---|---|---|---|---|
| Flt2 | 26/07/2013 | 57.5° N, 0.2° W | FL360 | 230s |
| Flt4 | 31/07/2013 | 56.7° N, 1.7° E | FL250 | 1260s |
| Flt9 | 08/08/2013 | 47.6° N, 6.5° E | FL320 | 480s |

As outlined in Sect. 3 D., the lidar backscatter signal is stored shot-by-shot with a range resolution of 5 m. In a first offline low-level data reduction procedure, the signal mean before the laser pulse is used for determination of the background light (i.e. sun) level and subtracted. Further, the GPS time stamp (arriving with one pulse-per-second PPS) is interpolated for each lidar signal with ~ 1 µs accuracy.

Since we are interested in relative fluctuations of the (molecular) backscatter coefficient (according to Eq. 3) only, the data analysis is straight-forward what may be illustrated by the familiar lidar equation:

$$S_{LI}(R) = \eta_L \cdot E_p \cdot c/2 \cdot A_L \cdot 1/R^2 \cdot O(R) \cdot \beta(R) \cdot T_{atm} \tag{7}$$

Following corrections are applied: Each lidar signal is multiplied by $R^2$. As worked out in Sect. 4 B., we may assume that the backscatter and thus the extinction coefficients from aerosols are negligible. Thus the atmospheric transmission $T_{atm} = exp\left(-2\int_0^R \alpha(R)\right)$ was estimated from local temperature and pressure measurements of the aircraft air data system. The laser pulse energy correction factor $E_p$ corrects for shot-to-shot fluctuation. The lidar overall efficiency $\eta_L$, speed of light $c$ and collecting telescope area $A_L$ are constants. The individual lidar signals are normalized to unity by applying a global overlap function determined from the whole dataset.

The lidar signal variance over the range, valid for the entire dataset, is then determined with the time variance method as will be detailed in the next section. Fig. 14 gives the *SNR* derived from the variance for the three flight legs.

For an evaluation, a simulation is performed based on the lidar equation with the respective atmospheric conditions of the flights and background, detection and photon noise. From the measured dispersion we infer that the collected signal is by a factor of approximately five lower than expected (in the simulation of Fig. 14, this is respected by an additional loss factor). This is complemented by the observation of a moderate slope overlap curve nearly asymptotically approaching unity. Hence, we suspect an optical misalignment in the receiver, such as the field stop or collimation, a matter that could not be carefully checked, aligned and optimized on ground due to the safety restrictions at Schiphol airport.

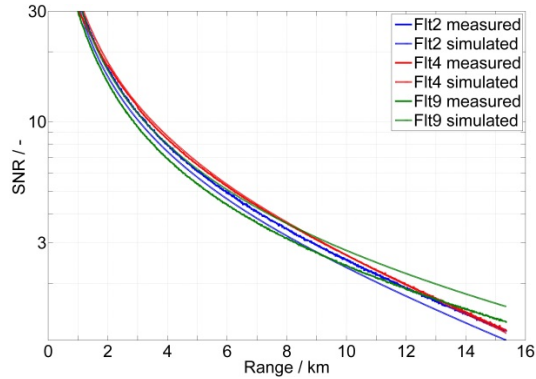

Fig. 14. *SNR* comparison of the three 'zero' measurements to simulation (thick lines: measurements, dotted: simulation).

The other slight deviations of the observed *SNR* from the predicted ones are mainly due to assumptions on the solar background level, assumed 300 W/m²·sr·μm for Flt2 and Flt4 (daytime) and 10 W/m²·sr·μm for Flt9 (night).

Despite the signal loss due to a misalignment, the SNR shows expected behavior with an approximate linear relationship between dispersion (standard deviation) and range.

### D. Detection noise and averaging

We should recall from the Introduction that the key to remotely detect CAT with the present method is to resolve tiny density, thus backscatter fluctuations on the (sub-)percent level. This goal may only be achieved by a substantial averaging of subsequent lidar acquisitions. In order to reduce the dispersion of the mean of a set of measurements with the square root of the number of averaged measurements, the respective measurements have to be afflicted with pure white noise only. This section aims at demonstrating the favorable noise characteristics of the lidar system.

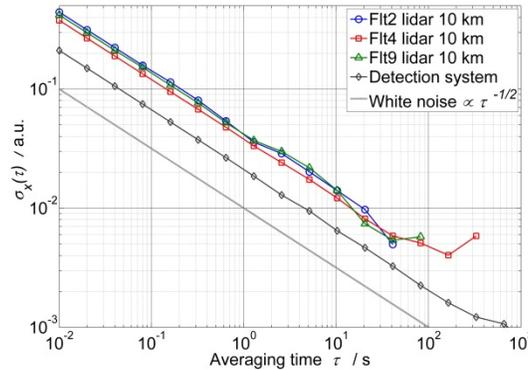

Fig. 15. Lidar measurement (at 10 km) averaging evaluated with time variance (time deviation shown here). The lidar measurements show white noise behavior still above 100 s integration time. The detection system alone has an even better stability (black dots, in arbitrary units). For illustration, the plot shows the white noise slope with $N_\Sigma^{-1/2}$ (grey line).

For this evaluation we employ the "Time Variance" TVAR or $\sigma_x^2(\tau)$, which typically is used for phase (time) data, but may equally well used for amplitude data. TVAR is a normalized variant of the "Modified Allan Variance" MVAR, matched to the classical variance. MVAR has the advantage (w.r.t. the original "Allan Variance" AVAR) to distinguish white phase (in our case amplitude) noise from flicker (1/f) noise. TVAR may be used as a robust and conservative estimator for the dispersion (variance) of an average, being valid for the whole analyzed dataset. Since any technical (in particular electronic) system is afflicted with flicker noise, MVAR or TVAR illustrate the maximum useful averaging time (or number), before the flicker floor sets in.

We employ TVAR over the averaging times $\tau = N_\Sigma \cdot \tau_0 = N_\Sigma / f_{rep}$ at different detection ranges. Fig. 15 shows the thus determined dispersion of the averaged data set for different averaged numbers $N_\Sigma$ for the three flight legs and the lidar measurement at 10 km distance. For a single measurement ($\tau_0 = 0.01$ s) the plot starts with the standard deviations for the single measurement as may be determined from the SNR in Fig. 14 (at 10 km). It then decreases with a $\tau^{-1/2}$ slope until reaching the flicker floor after 10,000 averaged measurements or 100 s integration time. Here, the values of Flt4 provide the highest confidence, and the longest evaluation period, due to the length of the dataset (cf. Table 6).

The figure also gives the noise behavior of the detection system alone, without laser pulse (acquired during Flt4 at the same time, just before laser pulse emission, actually providing the background level. Units are arbitrary, though.).

This examination shows that the lidar instrument already fulfils the formulated project requirements, the density (backscatter) measurement being resolvable on the percent level at 5 km distance after 3 s integration time (see Fig. 16).

On the other hand, integration times of some tens of seconds may be regarded as of mere academic nature considering the present application of a forward-pointing remote sensing instrument in a jet plane at a speed of 160 m/s (for DELICAT's PH-LAB) or 250 m/s in an application on a commercial airliner. Therefore, a spatial averaging should be performed as well. For the present lidar detection system, a 30 MHz sampling rate has been implemented which may be regarded as a maximum, shorter range gates not being necessary for turbulence resolution. The variance analysis has also been performed on synthesized longer range gates of 50 m, 100 m and 500 m which is shown in Fig. 16 for Flt4 (longest dataset) at 5 km distance.

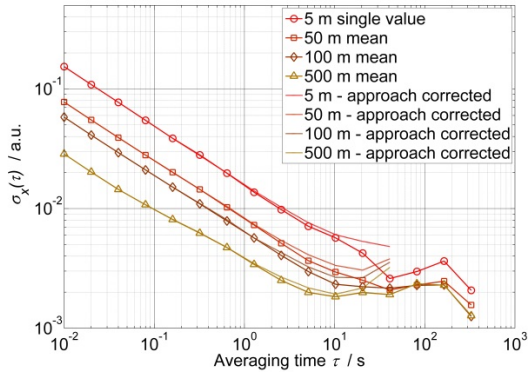

Fig. 16. Lidar measurement (at 5 km) averaging performed spatially and temporally. The light lines show the effect of approaching the target distance.

The analysis shows that by spatial averaging, a noise floor of 2–3 ‰ for the density measurement may be reached after some seconds integration time (the variance values at 300 s averaging time being both of low confidence and not applicable). As we will see in the next section, this low level is more than sufficient to resolve density fluctuations engendered by 'realistic' turbulence.

When approaching a certain target distance ('detection' distance, i.e.) with a simultaneous averaging (over some seconds, say), one incorporates measurements with higher dispersion (cf. Fig. 14) since performed farther away. This degrades the final dispersion of the averaged set. This effect is also shown in Fig. 16, with the light lines. It actually becomes only important for significant averaging times, for shorter ones the effect is merely noticeable. We should however infer, for a second time, that averaging times over longer times than ≈ 10 s are not suitable for this application.

One further observes that the spatial averaging, i.e. the range gate combining, does not fully obey the Poisson-noise rule of $\sigma_{N_\Sigma} \propto 1/\sqrt{N_\Sigma}$ which is illustrated in Fig. 17.

This behavior is not surprising since contiguous range gate measurements must be highly correlated due to the 100 ns pulse response of the amplification system (cf. Sect. 3.C). On the contrary, it is rather pleasant that it is so close to white noise (factor < 2). For a targeted observation of turbulence one should thus implement an optimized sampling rate and detection chain. The merit of such an optimisation of the detection system range gate is shown in the lower panel of Fig. 17.

In this section the noise behavior of the DELICAT lidar system has been analyzed in order to highlight its capability to determine the subtle air density fluctuations provoked by clear air turbulence. For adequate averaging over time and range, the system adheres to white (Poisson) noise allowing an improvement of the measurement set with factors of more than 250, depending on averaging time and range interval.

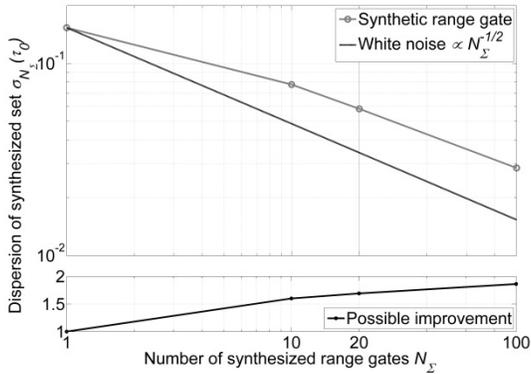

Fig. 17. Decrease of the dispersion of the combined 'synthesized' range gates (here at $\tau_0$), compared to Poisson noise. The lower panel shows the quotient.

### E. Discussion and instrument scaling

The previous section has shown that the present DELICAT lidar system is able to resolve the targeted density fluctuation by averaging the Rayleigh backscatter signal or directly the variance as proposed by [5]. In particular, the developed lidar attained and surpassed the requirements formulated within the DELICAT project agreement as mentioned in Sect. 2.

During the flight campaign 2013 of the DELICAT project, such turbulent events (and stronger) have been targeted. Though, during the > 30 flight hours, only few noticeable CAT events have been encountered, all merely of the 'light' category according to the flight crew. Vertical accelerations of the aircraft remained below 0.1 $g$ rms with very few excursions to 0.5 $g$ peak values ($g$ being the gravity acceleration). An analysis of the fast total air temperature probe showed that the temperature (and thus density] fluctuations did not exceed the 1.5 ‰ level for the strongest event and with an order of magnitude less for other light turbulent events. All recorded events thus remained well below the above shown sensitivity driven to its maximum. This meager yield in CAT encounter was, as noted above, mainly due the general meteorological conditions over Europe and the North-Eastern Atlantic during the campaign time frame. Another cause is certainly the very physical nature of CAT being spatio-temporally very constraint.

The principal ability of Rayleigh lidar to detect these subtle density variations has been shown by ground lidar [5]. From the airborne campaign we may draw the conclusion that the developed airborne lidar would have been capable to detect CAT with properties as described in the following.

In aviation, CAT is preferably classified by the cubic root of the kinetic energy dissipation rate $EDR = \sqrt[3]{\varepsilon}$, as originally suggested already by [35] and supported and operationally used by ICAO. It is matched to pilot-perceived levels as follows:

**Table 5: Turbulence levels according to [36]**

| $EDR$ / m$^{2/3}$/s | Level |
|---|---|
| < 0.1 | none |
| 0.1 – 0.4 | light |
| 0.4 – 0.7 | moderate |
| > 0.7 | severe |

From this turbulence severity classification scheme, vertical gust velocities $\sigma_w$ may be derived with the relationship [37, 38]:

$$\varepsilon^{2/3} = \frac{55}{9}\left(\frac{\Gamma(\frac{5}{6})}{\Gamma(\frac{1}{3})}\right)^{5/3} \frac{1}{\pi^{1/6}} \cdot \frac{1}{\alpha} \cdot \frac{1}{L_i^{2/3}} \sigma_w^2 \cong 0.78 \cdot \frac{1}{L_i^{2/3}} \sigma_w^2 \qquad (8)$$

where $\alpha^{-3/2} = C_1$ is the Kolmogorov-constant. Here, the value $C_1 = 0.53$ as suggested by [39] and [40] is used. $L_i$ is a scale length of the inertial subrange of the turbulence; throughout the aeronautics and aerodynamics domain, the longitudinal length scale $L_i = L_u = 2{,}500$ ft is suggested [41, Annex A, p. 682].

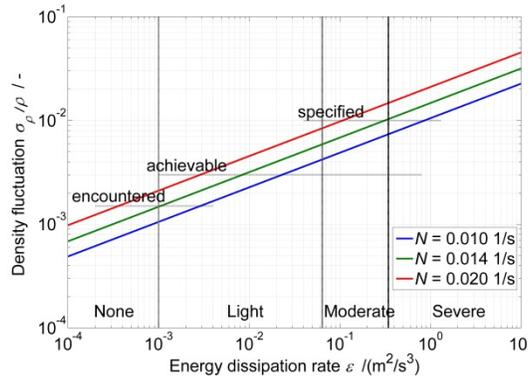

Fig. 18. Expected density fluctuation derived from turbulence severity as stipulated by ICAO, given for different values of Brunt-Väisälä-frequency. The figure also shows the different levels of lidar resolution, as specified by the DELICAT project, as shown above as achievable by appropriate averaging, and as encountered during the test flights.

With Eq. (4) one may thus derive the expected air density fluctuation $\sigma_\rho$:

$$\frac{\sigma_\rho}{\rho} = \sigma_w \frac{N}{g}$$

This relationship is shown in Fig. 18, with three representative values of the static stability $N$, for the high troposphere, the UTLS, and the lower stratosphere [42, e.g.].

Within the DELICAT project, so called 'moderate or greater' turbulences (MOG) were aimed for, considering the final application to aircraft safety. More specifically, a 1 % density fluctuation (detected at 5 km distance) was postulated which visibly relaxed the requirement on the lidar, but exacerbated the requirement on the strength of the turbulence to be encountered. Fig. 15 showed us that the flicker noise floor of the lidar detection system is not yet reached – with an indication of a possible roll-off at around 1 ‰ meaning that with less initial measurement dispersion, an even higher resolution may be reached technically. On the other hand, it is apparent from the above plot (Fig. 18) that the built lidar - as is - should prove sufficient for the aimed application of MOG and even light-to-moderate turbulence.

For a future airborne validation of this turbulence lidar, we may thus rather aim to increase the detection range. A last stability plot will show the hypothetic performance based on the evaluated measurements of Flt4 at 15 km distance. These measured data may be approximated by:

$$\sigma_x^2(\tau) = K_1^2 \cdot \frac{1}{\tau} + K_2^2 \cdot \tau \qquad (9)$$

with $K_1 = 0.0755 \text{ s}^{1/2}$ and $K_2 = 0.001 \text{ s}^{-1/2}$. These measured data are scaled with the following characteristics:
- Proper alignment, thus 5-fold gain in photon number – scale factor $1/\sqrt{5}$ (see Sect. 4.C);
- Optimized native range resolution to 50 m with synthesized to 100 m as per Fig. 17 – scale factors $1/\sqrt{10}$ and $1/1.3$ . Such range gate values have also been suggested in [4];
- Increase of maximum detection range to 25 km – scale factor $25^2 / 15^2$ (neglecting extinction).

This hypothetical performance is plotted in Fig. 19 showing that approximately the needed resolution will be reached in order to remotely (at 25 km distance) detect light-to-moderate and MOG turbulence.

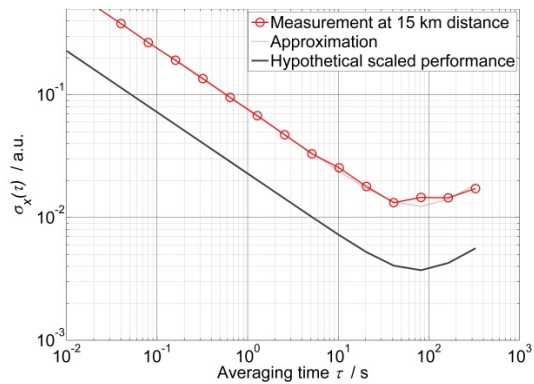

Fig. 19. Hypothetical performance of the present, slightly optimized lidar system for operation at 25 km distance.

A 25 km detection distance would mean a 2.5 min travel time for the actual PH–LAB; but it would mean a 100 s travel/warning time for a typical airliner (flying at Ma = 0.8). This would allow, for instance, up to some seconds integration (averaging) time, some hundred ms calculation, giving the flight crew a minute for a decision and once 'seat belt sign' given, the passengers and cabin crew 30 s for reaching their seats. This short exercise should show that even with such experimental scientific hardware, realistic turbulence protection may be achieved.

For a hypothetic future product further ameliorations should become available, as there are: Slightly larger receiver telescope aperture, optimized receiving layout as off-axis telescope for obstruction ratio minimization, more efficient laser and harmonic generation, more efficient detector. On the other side of the balance there should be a device for rejection of the aerosol-contaminated central part of the backscattered spectrum, or a differently mannered distinguishing of the broad molecular and narrow aerosol spectral components, as described in [43]. Such an interferometer-based device may decrease the effective signal by a factor two to five.

In the meantime, though, the optimized scientific lidar should validate the density fluctuation approach by collecting data on turbulence of significant level.

## 5. CONCLUSION AND OUTLOOK

Within the European DELICAT project, an airborne lidar system for the remote detection of CAT was designed and built by DLR and partners. A dedicated flight campaign on the Dutch research aircraft PH-LAB allowed acquiring reference data in different flight altitudes and conditions. Here, we analyzed signal, noise and stability characteristics that are a prerequisite for resolving the subtle air density fluctuations caused by clear air turbulence. The airborne lidar showed a factor of three better resolution than the project requirement, thus principally allowing the remote detection of CAT of light-to-moderate severity ($EDR \geq 0.2 \text{ m}^{2/3}/\text{s}$). Further analysis showed that the present system may be tuned and optimized, with acceptable effort, in order to extend the effective detection range up to 25 km. This would be within the range of the real application in aeronautics safety. Such technology may then be matured for improving flight safety by mitigation of CAT encounters.

On more general level, a versatile and modular airborne lidar in forward-pointing configuration was validated, opening the way for other aeronautics-related lidar-based detection schemes, such as gust detection by Doppler wind lidar (DWL). A respective receiver sub-assembly for near-range DWL measurement by fringe-imaging with a field-widened Michelson interferometer is in preparation right now [15]. Further applications cover the remote identification of crystalline icing conditions or areas contaminated with hazardous levels of volcanic ash or mineral dust. The combination of this versatile lidar on the aircraft platform allows for testing and verification of the respective detection techniques.

### Ackknowledgement

The research leading to these results has received funding from the European Union Seventh Framework Programme FP7/2007-2013 under grant agreement n° 233801 (DELICAT).

___